\documentclass{article}

\usepackage{arxiv}

\usepackage[utf8]{inputenc} 
\usepackage[T1]{fontenc}    
\usepackage{url}            
\usepackage{booktabs}       
\usepackage{amsfonts}       
\usepackage{nicefrac}       
\usepackage{microtype}      
\usepackage{lipsum}		
\usepackage{graphicx}
\usepackage{natbib}
\usepackage{doi}

\title{A Decision Framework for Blockchain Adoption}

\author{ \href{https://orcid.org/0000-0002-0907-3252}{\includegraphics[scale=0.06]{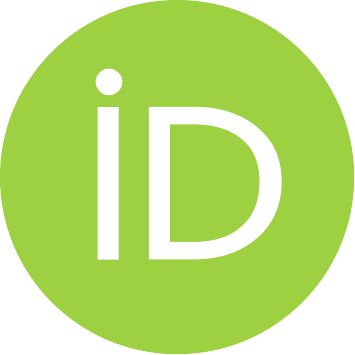}\hspace{1mm}Vittorio Capocasale} \\
	\texttt{vittorio.capocasale@polito.it} \\
	\And
	\href{https://orcid.org/0000-0001-6900-9917}{\includegraphics[scale=0.06]{orcid.pdf}\hspace{1mm}Guido Perboli} \\
	\texttt{guido.perboli@polito.it} \\
}

\renewcommand{\shorttitle}{A Decision Framework for Blockchain Adoption}

\begin{document}
\maketitle

\begin{abstract}
Blockchain and distributed ledger technologies are gaining the interest of the academy, companies, and institutions. Nonetheless, the path toward blockchain adoption is not straightforward, as blockchain is a complex technology that requires revisiting the standard way of addressing problems and tackling them from a decentralized perspective. Thus, decision-makers adopt blockchain technology for the wrong reasons or prefer it to more suitable ones. This work presents a decision framework for blockchain adoption to help decision-makers decide whether blockchain is applicable, valuable, and preferable to other technologies. In particular, The decision framework is composed of a small set of questions that can be answered from a managerial standpoint and that do not require a deep technical knowledge of blockchain-related topics.
\end{abstract}


\section{Introduction}

At present, companies are undergoing radical transformations based on information sharing and digitization, known collectively as the Industry 4.0 revolution \citep{Fakhri20201175}. Such a revolution is driven by the recent technological advancements in physical monitoring, data elaboration, virtualization, and automation technologies \citep{Bai2020, Boccia2021131, Caselli22Mixed, Fadda20211291}. On one side, data acquisition and storage is becoming cheaper and more accurate \citep{Perboli2021}; on the other, peer-to-peer technologies such as blockchain\citep{HE2022} and the Interplanetary File System \citep{Capocasale20221684} are transforming existing business paradigms \citep{Perboli201862018}. 

Blockchain creates trust among non-trusting parties without relying on intermediaries. A blockchain is composed of a network of nodes managing a shared and distributed database: tampering attempts are prevented by replicating the state of the database on each node. Smart contracts are independently executed by each node and are used to alter the state of the database. Thus, by leveraging the tamper-resiliency of the blockchain, smart contracts could enhance the fairness of critical processes, protect valuable resources, and automate business operations. Given the relevance of such topics \citep{Guastalla2022, serrano2022verification}, smart contract-based alternatives to existing services are surging in multiple sectors, including finance \citep{Pavlova2020687}, insurance \citep{Gatteschi2018}, logistics\citep{Pan20212079}, energy \citep{ruffini2022net, Khan2021}, and more. 

Nonetheless, blockchain is a complex technology that introduces many compromises and issues at all technical, legal, and economic levels. Thus, decision-makers often lack the necessary knowledge to make informed decisions on blockchain adoption, and misconceptions are widespread in the field \citep{Schneider2022}. Unsurprisingly, blockchain is often chosen for the wrong reasons or is preferred to better technologies \citep{Belotti20193796, Halaburda201827, Carson2018118, Labazova20194555}. Consequently, many blockchain projects do not last long or fail to fulfill the original goals \citep{Kaufman2021Consortium}. In this context, it is essential to develop standards and tools to simplify the managerial decision process on blockchain adoption.

We contribute to the current body of knowledge by making the following contributions:
\begin{itemize}
    \item we propose a decision flowchart for blockchain adoption that helps decision-makers understand when blockchain is applicable, valuable, and preferable to other solutions. The framework does not require any deep technical knowledge of blockchain technology and can be effectively employed by decision-makers with different backgrounds;
    \item we discuss the rationale behind each of the decision drivers of our framework to shed some light on some of the hidden caveats of blockchain technology, as they are not sufficiently discussed in the existing literature.
\end{itemize}

The remaining part of this study is structured as follows: Section \ref{sec:background} briefly describes the blockchain technology and presents a summary of the related works; Section \ref{sec:main} describes the blockchain adoption decision framework; Section \ref{sec:conclusion} concludes the study.

\section{Background}\label{sec:background}
This section summarizes the main concepts related to blockchain. Moreover, this section includes a summary of the related works. 
\subsection{Blockchain}

Blockchain is a technology that enables data sharing among non-trusting parties. Blockchain allows for solving trust issues without leveraging trusted third parties. Blockchain is composed of a network of peers that share a common database. The shared database is a ledger, as data can only be appended to it. Each peer manages its copy of the ledger independently from the others. Thus, peers can maliciously alter their copy, but not the global state of the ledger, which is decided based on what is stored in the majority of the copies \citep{Zheng2018352, Luo2022956}. We assume a uniform distribution of voting power among the peers to simplify the discussion. However, when we refer to the majority of the peers, we mean the majority of the voting power. 

Blockchains can be categorized according to their governance model as follows \citep{buterinPublicPrivate, Lin2017653}.
\begin{itemize}
 \item \textbf{Public}---Any peer can join the blockchain system and gain voting power. Public blockchains solve trust issues among their participants, as peers can autonomously validate transactions.
 \item \textbf{Consortium}---The blockchain system is managed by some well-identified peers who can set the rules for interacting with the ledger and gaining voting power. Consortium blockchains solve trust issues among the consortium members.
 \item \textbf{Fully private}---A single peer manages the blockchain system. Thus, the system is not decentralized.
\end{itemize}

\subsection{Problem Statement}
Many real-world systems are intrinsically decentralized. For example, supply chains are composed of numerous companies, and the behavior of each one affects the performance of the whole supply chain. Thus, it is logical to manage supply chains in a decentralized way by allowing each company to vote on the best strategy to improve the performance of the whole supply chain.

The introduction of blockchain technologies has created the opportunity to decentralize the management of data. Thus, blockchain has gained adoption in all those intrinsically decentralized systems that previously relied on trusted third parties. Nonetheless, blockchain is a complex technology and introduces many hidden compromises and issues. Moreover, decision-makers often lack the necessary technical knowledge to make informed decisions on blockchain adoption. Thus, blockchain is often adopted for the wrong reasons or is preferred to better technologies \citep{Belotti20193796, Halaburda201827, Carson2018118, Labazova20194555}.

To solve this problem, we created a framework that helps decision-makers understand when blockchain is applicable, valuable, and preferable to other solutions. The framework does not require a deep technical knowledge of blockchain technology and can be effectively employed by decision-makers with different backgrounds.

\subsection{Related Works} \label{sec:literature} 
According to Ref. \citep{Almeshal2021155425}, blockchain suitability frameworks can be divided into three categories: decision models, conceptual frameworks, and decision flowcharts.

Decision models use mathematical models to decide on blockchain adoption. For example, BAF is a framework for determining the ideal blockchain solution based on a weighted evaluation of detailed user requirements \citep{Gourisetti20201142}. 

Conceptual frameworks identify the factors to consider in adopting blockchain technologies based on the practical experience of researchers. For example, ten technology-driven factors are considered by the framework proposed in Ref. \citep{Scriber201870}. Other authors included non-technological factors (e.g., environmental considerations, such as regulations) \citep{Clohessy2020501, Labazova2019}. Open-ended questions that should be addressed when considering blockchain adoption are proposed in Ref. \citep{Angelis2019307}.

Decision flowcharts are based on graphs where nodes represent closed-ended questions and edges represent the related answers. Users are led to a decision by the path dictated by the answers they pick. Many authors adopted such a strategy in the literature. A study proposed a multi-step framework to decide on blockchain adoption and the type of blockchain needed \citep{Peck201738}. Others deepened the discussion by providing some guidelines on implementing working solutions \citep{Belotti20193796}, considering the security threats of using blockchain \citep{Puthal202153}, and analyzing real-world use cases \citep{Hassija2021, Wust201845, Gatteschi201862}. A framework explicitly designed for managers is proposed in Ref. \citep{challener2019blockchain}. Multiple frameworks were analyzed and condensed into a single one in Ref. \citep{koens2018blockchain}. However, we do not agree with some of the decision drivers proposed in all such works. For example, requiring the presence of multiple writers is unnecessary: a group of entities may need to record in a tamper-proof way what is written by a third one. In such a case, a blockchain could be a viable solution, as the multiple record keepers could prevent the writer from altering past data. Thus, blockchain adoption should be driven by the presence of multiple decision makers (the keepers can decide which data are alterable), not multiple writers.

Ref. \citep{Lo2018158} describes a framework composed of seven main questions and four subquestions. However, a good technical understanding of the technology is necessary to answer some of the proposed questions. A ten-step decision framework is proposed in Ref. \citep{Pedersen201999}. The framework is very useful as it considers many aspects that are ignored in similar works.

Finally, some authors proposed decision frameworks for blockchain adoption that are tailored to specific use cases (e.g., logistics \citep{Ar2020, ganeriwalla2018does, Hribernik2020} and the construction industry \citep{Hunhevicz2020}).

\begin{table*}[htpb] 
\caption{The table resumes the literature on decision frameworks for blockchain adoption based on decision flowcharts. The table highlights the common adoption questions between our framework and those proposed in the literature}
\centering


\begin{tabular}{|c|c|c|c|c|c|c|c|c|c|c|c|} \hline
Ref. & Q1 & Q2 & Q3 & Q4 & Q5 & Q6 & Q7 & Q8 & Q9 & Q10 & Q11 \\\hline
\citep{Hunhevicz2020}           & No  & Yes & No  & No  & No  & Yes & No  & No  & No  & No  & No  \\\hline
\citep{Wust201845}              & No  & Yes & No  & No  & No  & No  & No  & No  & No  & No  & No  \\\hline
\citep{Belotti20193796}         & No  & Yes & No  & No  & Yes & No  & No  & No  & No  & No  & No  \\\hline
\citep{Puthal202153}            & No  & Yes & No  & No  & Yes & No  & No  & No  & No  & Yes & Yes \\\hline
\citep{Hassija2021}             & No  & Yes & No  & No  & Yes & No  & No  & No  & No  & No  & No  \\\hline
\citep{Pedersen201999}          & Yes & Yes & No  & No  & Yes & No  & Yes & No  & No  & Yes & No  \\\hline
\citep{koens2018blockchain}     & No  & Yes & No  & No  & No  & No  & No  & No  & No  & No  & No  \\\hline
\citep{Peck201738}              & No  & Yes & No  & No  & No  & No  & No  & No  & No  & Yes & No  \\\hline
\citep{Gatteschi201862}         & No  & Yes & No  & No  & Yes & No  & No  & No  & No  & Yes & No  \\\hline
\citep{Lo2018158}               & Yes & Yes & No  & No  & Yes & No  & No  & No  & No  & Yes & No  \\\hline
\citep{challener2019blockchain} & Yes & No  & No  & No  & No  & Yes & No  & Yes & No  & Yes & No  \\\hline
This work                      & Yes & Yes & Yes & Yes & Yes & Yes & Yes & Yes & Yes & Yes & Yes \\\hline

\end{tabular}
\label{tab:summary}
\end{table*}

\section{Blockchain Adoption Decision Framework} \label{sec:main}
\begin{figure}[ht]
    \centering
    \includegraphics[scale=0.85]{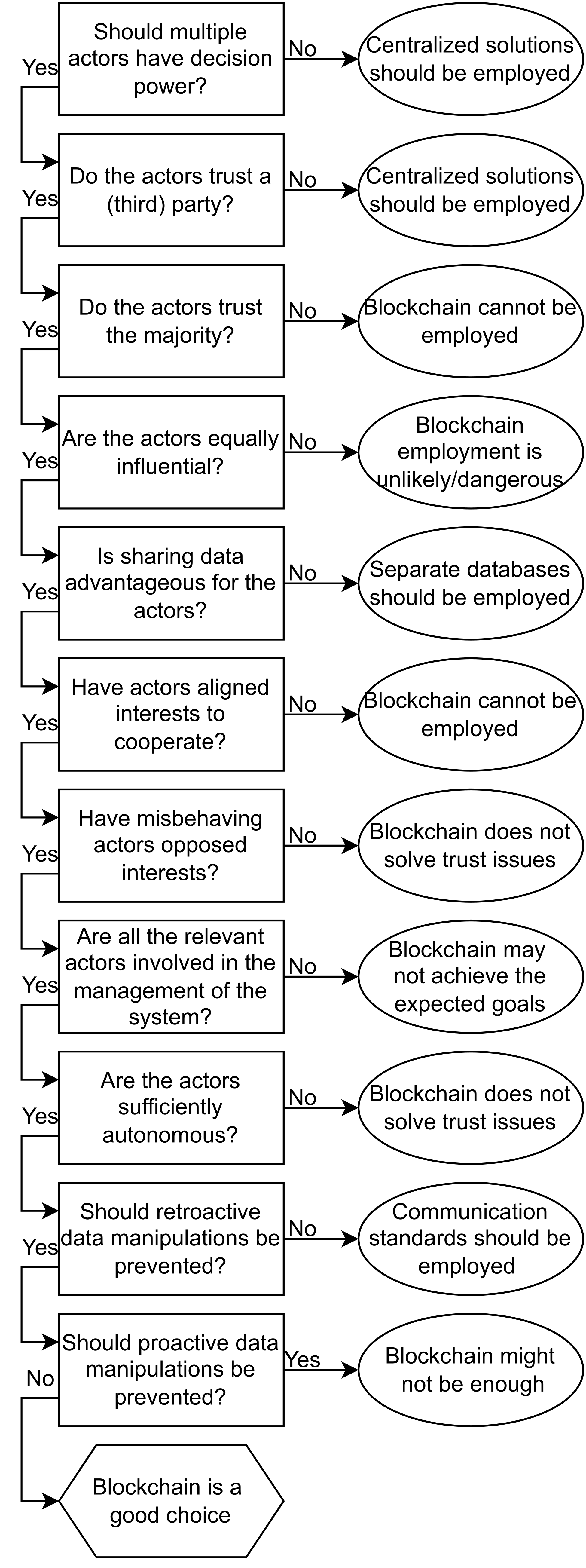}
    \caption{The proposed decision framework for blockchain adoption}
    \label{fig:flow}
\end{figure}

This section describes our decision framework for blockchain adoption, which is graphically summarized in Fig. \ref{fig:flow}.
Before introducing our approach, we want to make a few important remarks.
\begin{itemize}
    \item Blockchain is meaningful when decentralized governance is required. Even though the locution distributed ledger technology has gained adoption, decentralization is what matters, not distribution \citep{cfb}.
    \item Blockchain is inefficient and should be used only when necessary. Blockchain is the only technology allowing for managing a database in a decentralized fashion. However, if the database can be managed by a single entity, other technological solutions are better \citep{challener2019blockchain}.   
\end{itemize}

As a consequence of the previous points, fully private blockchains have little to no use, in our opinion. They can be employed to prevent accidental data modifications, but non-distributed ledgers are more efficient (e.g., ImmuDB \citep{paik2020immudb}). Thus, employing fully private blockchains can be a good marketing strategy but not a good technological one. For example, central bank digital currencies \citep{benedetti2022pow} should not leverage blockchain if they are managed by a single entity (the central bank). Consequently, our framework deals with the suitability of public or consortium blockchains. 

\subsection{Q1: Should Multiple Actors Have Decision Power?}
The most important thing to consider when deciding on blockchain adoption is whether or not the system is decentralized. If decision power is not shared among multiple actors, centralized solutions (e.g., distributed databases) should be preferred: according to the scalability trilemma, decentralization comes at the cost of scalability or security \citep{schaaf2021analysis}. Thus, compared to the blockchain, centralized solutions are more scalable and secure.

It is important to underline that having decision power means having voting power to validate write attempts in the context of blockchain. Blockchain can be described as a database that can be altered through a majority-based voting scheme. Thus, blockchain allows multiple actors to vote and decide which are valid database modifications. Remarkably, blockchain does not offer the same guarantees on reading attempts, as a single malicious actor could leak the contents of the database.

Finally, we underline that validating writing attempts does not imply having writing rights. For example, in a trial, the judge decides what is admissible as evidence but cannot produce evidence. As discussed in Sec. \ref{sec:literature}, this remark is one of the main points of differentiation between our work and the existing literature.

\subsection{Q2: Do the Actors Trust a (Third) Party?}
If an external entity or one of the actors is particularly trustworthy, the actors may be comfortable delegating their decision power to such an entity. In such a case, centralized solutions managed by the trusted party are better alternatives to blockchain for the same reasons outlined in the previous section. Vice versa, blockchain is a viable solution when no single party is trusted by all the actors. Almost all the decision frameworks present in the literature include this decision driver, which remarks its importance.

\subsection{Q3: Do the Actors Trust the Majority?}
Blockchain does not solve trust issues completely: a precondition for using blockchain technology is that the majority of the actors are trustworthy \citep{CapocasaleSmart22}. Thus, blockchain should only be used if actors are unlikely to collude. Otherwise, the malicious majority could tamper with or rewrite the database (51\% attack) \citep{Hao2022278}. Unfortunately, the literature has not given enough weight to such a relevant decision driver. We strongly encourage decision-makers to carefully consider the likelihood of 51\% attacks before adopting blockchain, as such attacks are not uncommon \citep{btcg51, etc51}. Launching 51\% attacks is easier on small networks as it requires fewer actors to collude. Moreover, some blockchains may be attacked by an even lower percentage of colluding peers as a consequence of the voting protocol in use. Thus, particularly in consortium blockchains, the presence of a trustworthy (super)majority must be carefully checked.

\subsection{Q4: Are the Actors Equally Influential?}
Actors should detain a similar decision power so that blockchain may become a viable solution: if one of the actors has strong leverage against the others, such an actor is likely to enforce its own centralized solution. In such a scenario, blockchain is unlikely to be successfully adopted, as the influential actor has no reason to share the control of the database. Moreover, even if a blockchain were used, the influential actor could force others to align with its own decisions. Thus, the majority would probably be untrustworthy. 

Porter's five forces analysis \citep{porter2008five} is useful to determine the influence of the various actors. In particular, studying the bargaining power of customers and suppliers can be helpful in determining the balance of power among actors, which could provide insights into the applicability of blockchain technology. 

For example, Amazon \citep{ritala2014coopetition} manages one of the biggest marketplaces in the world. Sellers have many advantages in selling their products on Amazon, including increased visibility, international expansion, and storage and shipping services. Nonetheless, sellers need to abide by Amazon's policies. Such policies are not negotiable, as Amazon has the upper hand in terms of bargaining power. In such a scenario, a blockchain solution is unlikely to be adopted, as Amazon can force the sellers to rely on Amazon's managed database. Conversely, blockchain could be adopted for the creation of a unified marketplace between Amazon and Alibaba \citep{Havinga2016}, as they are both e-commerce giants with similar bargaining power.  

\subsection{Q5: Is data sharing advantageous for the actors?}
Blockchains are shared databases. Actors that are not interested in sharing their data or receiving others' data are unlikely to join a blockchain network. Data that is not meant to be shared should be stored in centralized databases, as the database manager retains total power over the stored data. Nonetheless, blockchain can be leveraged in some use cases that require non-straightforward data-sharing approaches. We identified a few of them: partial sharing, delayed sharing, conditional sharing, and proof sharing.

Partial sharing refers to the necessity of sharing data with only some of the actors. In such a case, the best approach is to create a separate blockchain involving only the selected receivers. However, for costs or practicality, actors may prefer to store encrypted data into a single blockchain involving all the actors, and share the decryption key with only the selected receivers. Thus, encrypted data are stored in a tamper-proof database and shared with all the actors, but only who know the decryption key can recover the original data. Different encryption/decryption keys can be used to disclose data to different actor subsets.

Delayed sharing refers to the possibility of sharing data in the future while guaranteeing that it is not altered in the meanwhile. For example, some countries disclose classified documents after a certain amount of time. By storing encrypted documents in a blockchain and successively disclosing the decryption key, it is possible to guarantee the authenticity and integrity of the documents at disclosure time.

Conditional sharing refers to the possibility of sharing data only if a certain event occurs. For example, a company may have some confidential data that have to be shared only in case of litigation. Similar to the delayed sharing case, it is possible to use blockchain to guarantee the authenticity and integrity of encrypted data and successively reveal the decryption key, if necessary. Interestingly, if the event never occurs, blockchain is used to store data that is never disclosed.

Proof sharing refers to the possibility of not sharing the data directly but a proof computed on the data (e.g., zero-knowledge proofs \citep{HE2022}) or a fingerprint of the data (e.g., the hash of the data \citep{zemler2019concepts}). There are many reasons for using such approaches, including guaranteeing the integrity of the original data and minimizing the disclosure of information. Interestingly, blockchain is still used to share data, even if not in its original form. 

\subsection{Q6: Have the Actors Aligned Interests to Cooperate?}
Blockchain systems are based on majority consensus, which can be reached if actors are incentivized to behave correctly according to some common rules. If actors do not have aligned interests, they are unlikely to join a blockchain network, and even if they did, they would be unlikely to follow the same rules. Thus, blockchain is applicable only if cooperating is advantageous for the actors. 

Often, public blockchains offer economic incentives to align the actors' goals and persuade them to behave correctly. In consortium blockchains, other forms of incentives are common. Often, such incentives come indirectly in the form of business opportunities and cost savings. In logistics, for example, sharing data can be beneficial for demand forecasting and paperwork reduction \citep{Perboli20201713}. Moreover, by tracking assets along the whole supply chain, it is possible to easily assign responsibilities to actors and reduce the risk of litigation \citep{Perboli20201713}. Thus, the actors of a supply chain may share a common interest that can enable and sustain long-term cooperation.    

\subsection{Q7: Have Misbehaving Actors Opposed Interests?}
Even if actors have a strong motivation to cooperate, they might have an even stronger motivation to cheat, which is particularly true when opportunities to make quick and easy gains arise. For this reason, it is important to ensure that misbehaving actors have conflicting goals so that one's gains would mean another one's loss. In this way, the risk of majority collusion attempts is minimized, as actors should behave against their own interests to corrupt the system.

We examine the example of Bitcoin \citep{lansky2017bitcoin}. Each Bitcoin holder has good motivation to misbehave and create new Bitcoins, as this would increase the holder's purchasing power. However, creating new Bitcoins inflates the existing supply, reducing the purchasing power of each Bitcoin. Consequently, Bitcoin holders want to create Bitcoin for themselves while preventing others from doing the same. Thus, collusion attempts are unlikely, as Bitcoin holders have conflicting interests when misbehaving.

We examine the example of a group of friends betting on the winner of a horse race. For simplicity, we assume that each friend picks a different horse. If the friends do not want to rely on trusted third parties, they could rely on a smart contract to collect the money in advance and then forward them to the winner: the verifiability and tamper-proof property of the blockchain would guarantee the correct handling of the bet. Thus, blockchain may seem a good solution. Unfortunately, in this scenario, the majority of the friends will lose the bet and will likely collude to take back their money instead of forwarding the prize to the winner. Given that blockchain decisions (including the behavior of smart contracts) are based on majority agreements, the winner of the bet will not receive the prize. Thus, blockchain should not be employed when cheating attempts favor the majority.

\subsection{Q8: Are All the Relevant Actors Involved in the Management of the System?}
By leveraging blockchain, decisions can be taken through majority voting instead of being delegated to a trusted third party. Nonetheless, a blockchain system is a third party for the actors that do not have voting power. Thus, blockchain does not offer any additional trustworthiness guarantees to them, and blockchain members should not expect entities to acknowledge the trustworthiness of third-party managed blockchain systems.

In logistics, for example, consortium blockchains are often used to facilitate the exchange of data among supply chain companies \citep{HE2022}. Final retail consumers, however, are rarely part of the consortium, as they lack the means, the technical knowledge, the time, the economic incentives, and the will to be involved in the consortium. To them, logistic blockchains are trusted third parties. Thus, supply chain companies should not join a blockchain system solely to increase data transparency for the final consumers, as consumers have no reason to trust the data stored in a blockchain more than the data provided by their retailer. Thus, blockchain systems should be used to create value for their participants, not external entities.

\subsection{Q9: Are the Actors Sufficiently Autonomous?}
The resiliency of blockchain is proportional to its decentralization. Thus, actors should be as autonomous and independent as possible to guarantee sufficient resilience to errors and tampering attempts. If too many actors needed to rely on others to code smart contracts, keep an updated copy of the ledger, participate in the voting process, and validate transactions, the blockchain would become supposedly decentralized but substantially centralized. In such a condition, blockchain does not offer any benefits over centralized systems but still imposes significant scalability drawbacks. Thus, blockchain should not be used if true decentralization cannot be guaranteed. In particular, decentralization cannot be improved by increasing the number of blockchain nodes managed by each actor, as distribution and decentralization are different concepts \citep{cfb}.  

\subsection{Q10: Should Retroactive Data Manipulations Be Prevented?}
If a blockchain system is sufficiently decentralized, data stored in a blockchain ledger can be considered tamper-proof. Thus, data cannot be manipulated after insertion, and updates are only possible by appending a newer version of the data to the ledger. Nonetheless, blockchain would keep both versions, allowing actors to track changes. Thus, blockchain is a good solution if it is important to prevent retroactive data manipulations. Nonetheless, if such a property is not relevant, standardizing the data exchange protocols among the actors is enough to share data efficiently. Thus, if preventing retroactive data manipulations is not a priority, each actor should manage its centralized database and use a standard data sharing protocol to exchange information with the other actors. Interestingly, blockchains are often adopted solely to enforce standardization \citep{notDecentralization}. 

\subsection{Q11: Should Proactive Data Manipulations Be Prevented?}
As previously discussed, blockchain can prevent retroactive data manipulations. Nonetheless, blockchain can rarely prevent proactive data manipulations (i.e., manipulations that happen before the data is stored in the blockchain). In particular, oracle data can hardly ever be verified and validated, as such data often convey information about the physical world \citep{Perboli2021}. In logistics, for example, the temperature of a frozen product at a given time is likely measured by the actor that is handling that product at that time. The other actors cannot measure such a temperature, as they lack physical possession of the product. Thus, they need to rely on and cannot verify the accuracy of the measure of the handling actor. In such a scenario, blockchain is prone to the garbage in, garbage out problem. Conversely, Bitcoin transactions are fully digital, and each peer can independently verify them: by keeping a registry of the balance of each Bitcoin holder, each peer can determine who has enough coins to spend. Nonetheless, very few use cases can be modeled without relying on oracles, which limits blockchain usefulness when proactive data manipulations must be prevented.

\section{Conclusion}\label{sec:conclusion}
Interest in blockchain technology is rising among all individuals, countries, and companies. Nonetheless, the path toward blockchain adoption is not straightforward, as blockchain is a complex technology that requires revisiting the standard way of addressing problems and tackling them from a decentralized perspective. Thus, decision-makers may adopt blockchain technology for the wrong reasons or prefer it to more suitable ones.

This work presented a decision flowchart that can help readers to decide on blockchain adoption. We considered various decision drivers that should help the readers understand whether blockchain is applicable, valuable, and preferable to other technologies. We believe that our framework can be particularly useful to decision-makers, as it includes many decision drivers that are overlooked by other similar works in the literature. Moreover, our framework can be employed without deep knowledge of blockchain, as the concepts are discussed from a high-level perspective. Thus, we believe that it can be effectively used by managers to drive blockchain adoption in their companies without wasting time studying the very articulate blockchain technology.

Our framework underlines the correlation between decentralization and the security of blockchain systems. The main takeaway is that blockchain should only be employed if sufficient decentralization can be guaranteed, which is rarely the case.

Future work will be focused on extending the current framework to provide support for additional questions that may arise following blockchain adoption, including feature selection, platform binding, cost considerations, and more.

\bibliographystyle{unsrtnat}
\bibliography{template}

\begin{thebibliography}{60}
\providecommand{\natexlab}[1]{#1}
\providecommand{\url}[1]{\texttt{#1}}
\expandafter\ifx\csname urlstyle\endcsname\relax
  \providecommand{\doi}[1]{doi: #1}\else
  \providecommand{\doi}{doi: \begingroup \urlstyle{rm}\Url}\fi

\bibitem[Fakhri et~al.(2020)Fakhri, Mohammed, Khan, Sadiq, Alkazemi, Pillai,
  and Choi]{Fakhri20201175}
A.B. Fakhri, S.L. Mohammed, I.~Khan, A.S. Sadiq, B.~Alkazemi, P.~Pillai, and
  B.J. Choi.
\newblock Industry 4.0: Architecture and equipment revolution.
\newblock \emph{Computers, Materials and Continua}, 66\penalty0 (2):\penalty0
  1175--1194, 2020.
\newblock \doi{10.32604/cmc.2020.012587}.

\bibitem[Bai et~al.(2020)Bai, Dallasega, Orzes, and Sarkis]{Bai2020}
C.~Bai, P.~Dallasega, G.~Orzes, and J.~Sarkis.
\newblock Industry 4.0 technologies assessment: A sustainability perspective.
\newblock \emph{International Journal of Production Economics}, 229, 2020.
\newblock \doi{10.1016/j.ijpe.2020.107776}.

\bibitem[Boccia et~al.(2021)Boccia, Mancuso, Masone, and Sterle]{Boccia2021131}
M.~Boccia, A.~Mancuso, A.~Masone, and C.~Sterle.
\newblock A feature based solution approach for the flying sidekick traveling
  salesman problem.
\newblock \emph{Communications in Computer and Information Science}, 1476
  CCIS:\penalty0 131--146, 2021.
\newblock \doi{10.1007/978-3-030-86433-0\_9}.

\bibitem[Caselli et~al.(2022)Caselli, Delorme, Iori, and Magni]{Caselli22Mixed}
Giulia Caselli, Maxence Delorme, Manuel Iori, and Carlo~Alberto Magni.
\newblock Mixed integer linear programming for a real-world parallel machine
  scheduling problem with workforce and precedence constraints.
\newblock In Lavinia Amorosi, Paolo Dell'Olmo, and Isabella Lari, editors,
  \emph{Optimization in Artificial Intelligence and Data Sciences}, pages
  61--71, Cham, 2022. Springer International Publishing.
\newblock ISBN 978-3-030-95380-5.

\bibitem[Fadda et~al.(2021)Fadda, Fedorov, Perboli, and Barbosa]{Fadda20211291}
E.~Fadda, S.~Fedorov, G.~Perboli, and I.D.C. Barbosa.
\newblock Mixing machine learning and optimization for the tactical capacity
  planning in last-mile delivery.
\newblock In \emph{Proceedings - 2021 IEEE 45th Annual Computers, Software, and
  Applications Conference, COMPSAC 2021}, pages 1291--1296, 2021.
\newblock \doi{10.1109/COMPSAC51774.2021.00180}.

\bibitem[Capocasale et~al.(2021)Capocasale, Gotta, Musso, and
  Perboli]{Perboli2021}
Vittorio Capocasale, Danilo Gotta, Stefano Musso, and Guido Perboli.
\newblock A blockchain, 5g and iot-based transaction management system for
  smart logistics: an hyperledger framework.
\newblock In \emph{2021 IEEE 45th Annual Computers, Software, and Applications
  Conference (COMPSAC)}, pages 1285--1290. IEEE, 2021.

\bibitem[He et~al.(2022)He, Wang, Sun, Bie, Lan, Song, Zeng, Pustis{\u{e}}k,
  and Qiu]{HE2022}
Ming He, Haodi Wang, Yunchuan Sun, Rongfang Bie, Tian Lan, Qi~Song, Xi~Zeng,
  Matevz{\u{z}} Pustis{\u{e}}k, and Zhenyu Qiu.
\newblock T2l: A traceable and trustable consortium blockchain for logistics.
\newblock \emph{Digital Communications and Networks}, 2022.
\newblock \doi{https://doi.org/10.1016/j.dcan.2022.06.015}.

\bibitem[Capocasale et~al.(2022)Capocasale, Musso, and
  Perboli]{Capocasale20221684}
V.~Capocasale, S.~Musso, and G.~Perboli.
\newblock Interplanetary file system in logistic networks: a review.
\newblock In \emph{Proceedings - 2022 IEEE 46th Annual Computers, Software, and
  Applications Conference, COMPSAC 2022}, pages 1684--1689, 2022.
\newblock \doi{10.1109/COMPSAC54236.2022.00268}.

\bibitem[Perboli et~al.(2018)Perboli, Musso, and Rosano]{Perboli201862018}
G.~Perboli, S.~Musso, and M.~Rosano.
\newblock Blockchain in logistics and supply chain: A lean approach for
  designing real-world use cases.
\newblock \emph{IEEE Access}, 6:\penalty0 62018--62028, 2018.
\newblock \doi{10.1109/ACCESS.2018.2875782}.

\bibitem[Aringhieri et~al.(2022)Aringhieri, Bigharaz, Duma, and
  Guastalla]{Guastalla2022}
R.~Aringhieri, S.~Bigharaz, D.~Duma, and A.~Guastalla.
\newblock Fairness in ambulance routing for post disaster management.
\newblock \emph{Central European Journal of Operations Research}, 30\penalty0
  (1):\penalty0 189--211, 2022.

\bibitem[Serrano(2022)]{serrano2022verification}
Will Serrano.
\newblock Verification and validation for data marketplaces via a blockchain
  and smart contracts.
\newblock \emph{Blockchain: Research and Applications}, page 100100, 2022.

\bibitem[Pavlova(2020)]{Pavlova2020687}
I.~Pavlova.
\newblock Blockchain etfs: dynamic correlations and hedging capabilities.
\newblock \emph{Managerial Finance}, 47\penalty0 (5):\penalty0 687--702, 2020.
\newblock \doi{10.1108/MF-11-2019-0565}.

\bibitem[Gatteschi et~al.(2018{\natexlab{a}})Gatteschi, Lamberti, Demartini,
  Pranteda, and Santamaría]{Gatteschi2018}
V.~Gatteschi, F.~Lamberti, C.~Demartini, C.~Pranteda, and V.~Santamaría.
\newblock Blockchain and smart contracts for insurance: Is the technology
  mature enough?
\newblock \emph{Future Internet}, 10\penalty0 (2), 2018{\natexlab{a}}.
\newblock \doi{10.3390/fi10020020}.

\bibitem[Pan et~al.(2021)Pan, Zhou, Piramuthu, Giannikas, and
  Chen]{Pan20212079}
S.~Pan, W.~Zhou, S.~Piramuthu, V.~Giannikas, and C.~Chen.
\newblock Smart city for sustainable urban freight logistics.
\newblock \emph{International Journal of Production Research}, 59\penalty0
  (7):\penalty0 2079--2089, 2021.
\newblock \doi{10.1080/00207543.2021.1893970}.

\bibitem[Ruffini et~al.(2022)Ruffini, Salerno, and Sim{\~o}es]{ruffini2022net}
Alessandra Ruffini, Andrea Salerno, and Francisco Sim{\~o}es.
\newblock Net-zero emissions: main technological, geopolitical, and economic
  consequences of the new energy scenario.
\newblock \emph{Available at SSRN 3998525}, 2022.

\bibitem[Khan et~al.(2021)Khan, Laghari, Liu, Shaikh, Ma, Wang, and
  Wagan]{Khan2021}
A.A. Khan, A.A. Laghari, D.-S. Liu, A.A. Shaikh, D.-A. Ma, C.-Y. Wang, and A.A.
  Wagan.
\newblock Eps-ledger: Blockchain hyperledger sawtooth-enabled distributed power
  systems chain of operation and control node privacy and security.
\newblock \emph{Electronics (Switzerland)}, 10\penalty0 (19), 2021.
\newblock \doi{10.3390/electronics10192395}.

\bibitem[Schneider and Azan(2022)]{Schneider2022}
B.~Schneider and W.~Azan.
\newblock Perceptions and misconceptions of blockchain: The potential of
  applying threshold concept theory.
\newblock In \emph{2022 IEEE 6th International Conference on Logistics
  Operations Management, GOL 2022}, 2022.
\newblock \doi{10.1109/GOL53975.2022.9820452}.

\bibitem[Belotti et~al.(2019)Belotti, Božić, Pujolle, and
  Secci]{Belotti20193796}
M.~Belotti, N.~Božić, G.~Pujolle, and S.~Secci.
\newblock A vademecum on blockchain technologies: When, which, and how.
\newblock \emph{IEEE Communications Surveys and Tutorials}, 21\penalty0
  (4):\penalty0 3796--3838, 2019.
\newblock \doi{10.1109/COMST.2019.2928178}.

\bibitem[Halaburda(2018)]{Halaburda201827}
H.~Halaburda.
\newblock Economic and business dimensions blockchain revolution without the
  blockchain? most of the suggested benefits of blockchain technologies do not
  come from elements unique to the blockchain.
\newblock \emph{Communications of the ACM}, 61\penalty0 (7):\penalty0 27--29,
  2018.
\newblock \doi{10.1145/3225619}.

\bibitem[Carson et~al.(2018)Carson, Romanelli, Walsh, and
  Zhumaev]{Carson2018118}
B.~Carson, G.~Romanelli, P.~Walsh, and A.~Zhumaev.
\newblock Blockchain beyond the hype: What is the strategic business value?
\newblock \emph{McKinsey Quarterly}, 2018\penalty0 (4):\penalty0 118--127,
  2018.

\bibitem[Labazova et~al.(2019)Labazova, Dehling, and Sunyaev]{Labazova20194555}
O.~Labazova, T.~Dehling, and A.~Sunyaev.
\newblock From hype to reality: A taxonomy of blockchain applications.
\newblock In \emph{Proceedings of the Annual Hawaii International Conference on
  System Sciences}, volume 2019-January, pages 4555--4564, 2019.

\bibitem[Kaufman et~al.(2021)Kaufman, Heister, and
  Yuthas]{Kaufman2021Consortium}
Matt Kaufman, Stanton Heister, and Kristi Yuthas.
\newblock Consortium capabilities for enterprise blockchain success, 8 2021.
\newblock URL
  \url{https://jbba.scholasticahq.com/article/27744-consortium-capabilities-for-enterprise-blockchain-success}.

\bibitem[Zheng et~al.(2018)Zheng, Xie, Dai, Chen, and Wang]{Zheng2018352}
Z.~Zheng, S.~Xie, H.-N. Dai, X.~Chen, and H.~Wang.
\newblock Blockchain challenges and opportunities: A survey.
\newblock \emph{International Journal of Web and Grid Services}, 14\penalty0
  (4):\penalty0 352--375, 2018.
\newblock \doi{10.1504/IJWGS.2018.095647}.

\bibitem[Luo et~al.(2022)Luo, Hu, Zhang, Zhang, Liu, Diao, and
  Huang]{Luo2022956}
C.~Luo, Y.~Hu, S.~Zhang, Y.~Zhang, Y.~Liu, X.~Diao, and G.~Huang.
\newblock Fission: Autonomous, scalable sharding for iot blockchain.
\newblock In \emph{Proceedings - 2022 IEEE 46th Annual Computers, Software, and
  Applications Conference, COMPSAC 2022}, pages 956--965, 2022.
\newblock \doi{10.1109/COMPSAC54236.2022.00148}.

\bibitem[Buterin(2015)]{buterinPublicPrivate}
Vitalik Buterin.
\newblock On public and private blockchains, 2015.
\newblock URL
  \url{https://sawtooth.hyperledger.org/docs/core/releases/1.2.6/introduction.html}.

\bibitem[Lin and Liao(2017)]{Lin2017653}
I.-C. Lin and T.-C. Liao.
\newblock A survey of blockchain security issues and challenges.
\newblock \emph{International Journal of Network Security}, 19\penalty0
  (5):\penalty0 653--659, 2017.
\newblock \doi{10.6633/IJNS.201709.19(5).01}.

\bibitem[Almeshal and Alhogail(2021)]{Almeshal2021155425}
T.A. Almeshal and A.A. Alhogail.
\newblock Blockchain for businesses: A scoping review of suitability
  evaluations frameworks.
\newblock \emph{IEEE Access}, 9:\penalty0 155425--155442, 2021.
\newblock \doi{10.1109/ACCESS.2021.3128608}.

\bibitem[Gourisetti et~al.(2020)Gourisetti, Mylrea, and
  Patangia]{Gourisetti20201142}
S.N.G. Gourisetti, M.~Mylrea, and H.~Patangia.
\newblock Evaluation and demonstration of blockchain applicability framework.
\newblock \emph{IEEE Transactions on Engineering Management}, 67\penalty0
  (4):\penalty0 1142--1156, 2020.
\newblock \doi{10.1109/TEM.2019.2928280}.

\bibitem[Scriber(2018)]{Scriber201870}
B.A. Scriber.
\newblock A framework for determining blockchain applicability.
\newblock \emph{IEEE Software}, 35\penalty0 (4):\penalty0 70--77, 2018.
\newblock \doi{10.1109/MS.2018.2801552}.

\bibitem[Clohessy et~al.(2020)Clohessy, Treiblmaier, Acton, and
  Rogers]{Clohessy2020501}
T.~Clohessy, H.~Treiblmaier, T.~Acton, and N.~Rogers.
\newblock Antecedents of blockchain adoption: An integrative framework.
\newblock \emph{Strategic Change}, 29\penalty0 (5):\penalty0 501--515, 2020.
\newblock \doi{10.1002/jsc.2360}.

\bibitem[Labazova(2019)]{Labazova2019}
O.~Labazova.
\newblock Towards a framework for evaluation of blockchain implementations.
\newblock In \emph{40th International Conference on Information Systems, ICIS
  2019}, 2019.

\bibitem[Angelis and Ribeiro~da Silva(2019)]{Angelis2019307}
J.~Angelis and E.~Ribeiro~da Silva.
\newblock Blockchain adoption: A value driver perspective.
\newblock \emph{Business Horizons}, 62\penalty0 (3):\penalty0 307--314, 2019.
\newblock \doi{10.1016/j.bushor.2018.12.001}.

\bibitem[Peck(2017)]{Peck201738}
M.E. Peck.
\newblock Blockchain world - do you need a blockchain? this chart will tell you
  if the technology can solve your problem.
\newblock \emph{IEEE Spectrum}, 54\penalty0 (10):\penalty0 38--60, 2017.
\newblock \doi{10.1109/MSPEC.2017.8048838}.

\bibitem[Puthal et~al.(2021)Puthal, Mohanty, Kougianos, and Das]{Puthal202153}
D.~Puthal, S.P. Mohanty, E.~Kougianos, and G.~Das.
\newblock When do we need the blockchain?
\newblock \emph{IEEE Consumer Electronics Magazine}, 10\penalty0 (2):\penalty0
  53--56, 2021.
\newblock \doi{10.1109/MCE.2020.3015606}.

\bibitem[Hassija et~al.(2021)Hassija, Zeadally, Jain, Tahiliani, Chamola, and
  Gupta]{Hassija2021}
V.~Hassija, S.~Zeadally, I.~Jain, A.~Tahiliani, V.~Chamola, and S.~Gupta.
\newblock Framework for determining the suitability of blockchain: Criteria and
  issues to consider.
\newblock \emph{Transactions on Emerging Telecommunications Technologies},
  32\penalty0 (10), 2021.
\newblock \doi{10.1002/ett.4334}.

\bibitem[Wust and Gervais(2018)]{Wust201845}
K.~Wust and A.~Gervais.
\newblock Do you need a blockchain?
\newblock In \emph{Proceedings - 2018 Crypto Valley Conference on Blockchain
  Technology, CVCBT 2018}, pages 45--54, 2018.
\newblock \doi{10.1109/CVCBT.2018.00011}.

\bibitem[Gatteschi et~al.(2018{\natexlab{b}})Gatteschi, Lamberti, Demartini,
  Pranteda, and Santamaria]{Gatteschi201862}
V.~Gatteschi, F.~Lamberti, C.~Demartini, C.~Pranteda, and V.~Santamaria.
\newblock To blockchain or not to blockchain: That is the question.
\newblock \emph{IT Professional}, 20\penalty0 (2):\penalty0 62--74,
  2018{\natexlab{b}}.
\newblock \doi{10.1109/MITP.2018.021921652}.

\bibitem[Challener et~al.(2019)Challener, Vachino, Howard~II, Pikas, and
  John]{challener2019blockchain}
David~C Challener, Maria~E Vachino, James~P Howard~II, Christina~K Pikas, and
  Anil John.
\newblock Blockchain basics and suitability: A primer for program managers.
\newblock \emph{ournal of Information Technology Management}, 30\penalty0
  (3):\penalty0 33--44, 2019.

\bibitem[Koens and Poll(2018)]{koens2018blockchain}
Tommy Koens and Erik Poll.
\newblock What blockchain alternative do you need?
\newblock In \emph{Data Privacy Management, Cryptocurrencies and Blockchain
  Technology}, pages 113--129. 2018.

\bibitem[Lo et~al.(2018)Lo, Xu, Chiam, and Lu]{Lo2018158}
S.K. Lo, X.~Xu, Y.K. Chiam, and Q.~Lu.
\newblock Evaluating suitability of applying blockchain.
\newblock In \emph{Proceedings of the IEEE International Conference on
  Engineering of Complex Computer Systems, ICECCS}, volume 2017-November, pages
  158--161, 2018.
\newblock \doi{10.1109/ICECCS.2017.26}.

\bibitem[Pedersen et~al.(2019)Pedersen, Risius, and Beck]{Pedersen201999}
A.B. Pedersen, M.~Risius, and R.~Beck.
\newblock A ten-step decision path to determine when to use blockchain
  technologies.
\newblock \emph{MIS Quarterly Executive}, 18\penalty0 (2):\penalty0 99--115,
  2019.
\newblock \doi{10.17705/2msqe.00010}.

\bibitem[Ar et~al.(2020)Ar, Erol, Peker, Ozdemir, Medeni, and Medeni]{Ar2020}
I.M. Ar, I.~Erol, I.~Peker, A.I. Ozdemir, T.D. Medeni, and I.T. Medeni.
\newblock Evaluating the feasibility of blockchain in logistics operations: A
  decision framework.
\newblock \emph{Expert Systems with Applications}, 158, 2020.
\newblock \doi{10.1016/j.eswa.2020.113543}.

\bibitem[Ganeriwalla et~al.(2018)Ganeriwalla, Casey, Shrikrishna, Bender, and
  Gstettner]{ganeriwalla2018does}
Amit Ganeriwalla, Michael Casey, Prema Shrikrishna, Jan~Philipp Bender, and
  Stefan Gstettner.
\newblock Does your supply chain need a blockchain?
\newblock Technical report, The Boston Consulting Group, 2018.

\bibitem[Hribernik et~al.(2020)Hribernik, Zero, Kummer, and
  Herold]{Hribernik2020}
M.~Hribernik, K.~Zero, S.~Kummer, and D.M. Herold.
\newblock City logistics: Towards a blockchain decision framework for
  collaborative parcel deliveries in micro-hubs.
\newblock \emph{Transportation Research Interdisciplinary Perspectives}, 8,
  2020.
\newblock ISSN 25901982.
\newblock \doi{10.1016/j.trip.2020.100274}.

\bibitem[Hunhevicz and Hall(2020)]{Hunhevicz2020}
J.J. Hunhevicz and D.M. Hall.
\newblock Do you need a blockchain in construction? use case categories and
  decision framework for dlt design options.
\newblock \emph{Advanced Engineering Informatics}, 45, 2020.
\newblock \doi{10.1016/j.aei.2020.101094}.

\bibitem[{Come-from-Beyond}(2020)]{cfb}
{Come-from-Beyond}.
\newblock Decentralized vs distributed, or why dlt is (probably) an incorrect
  term, 2020.
\newblock URL
  \url{https://medium.com/@comefrombeyond/decentralized-vs-distributed-or-why-dlt-is-probably-an-incorrect-term-fccbf62bdfe7}.

\bibitem[Paik et~al.(2020)Paik, Iraz{\'a}bal, Zimmer, Meloni, and
  Padurean]{paik2020immudb}
Michael Paik, Jer{\'o}nimo Iraz{\'a}bal, Dennis Zimmer, Michele Meloni, and
  Valentin Padurean.
\newblock immudb: A lightweight, performant immutable database, 2020.
\newblock URL \url{https://arxiv.org/pdf/2207.06870.pdf}.

\bibitem[Benedetti et~al.(2022)Benedetti, De~Sclavis, Favorito, Galano,
  Giammusso, Muci, and Nardelli]{benedetti2022pow}
Marco Benedetti, Francesco De~Sclavis, Marco Favorito, Giuseppe Galano, Sara
  Giammusso, Antonio Muci, and Matteo Nardelli.
\newblock A pow-less bitcoin with certified byzantine consensus.
\newblock \emph{arXiv preprint arXiv:2207.06870}, 2022.

\bibitem[Schaaf et~al.(2021)Schaaf, Rezabek, and Kinkelin]{schaaf2021analysis}
Paul Schaaf, Filip Rezabek, and Holger Kinkelin.
\newblock Analysis of proof of stake flavors with regards to the scalability
  trilemma.
\newblock \emph{Network}, 63, 2021.

\bibitem[Capocasale and Perboli(2022)]{CapocasaleSmart22}
Vittorio Capocasale and Guido Perboli.
\newblock Standardizing smart contracts.
\newblock \emph{IEEE Access}, 10:\penalty0 91203--91212, 2022.
\newblock \doi{10.1109/ACCESS.2022.3202550}.

\bibitem[Hao(2022)]{Hao2022278}
Y.~Hao.
\newblock Research of the 51\% attack based on blockchain.
\newblock In \emph{2022 3rd International Conference on Computer Vision, Image
  and Deep Learning and International Conference on Computer Engineering and
  Applications, CVIDL and ICCEA 2022}, pages 278--283, 2022.
\newblock \doi{10.1109/CVIDLICCEA56201.2022.9824528}.

\bibitem[Martin(2020)]{btcg51}
Jack Martin.
\newblock Bitcoin gold blockchain hit by 51\% attack leading to \$70k double
  spend, 2020.
\newblock URL
  \url{https://cointelegraph.com/news/bitcoin-gold-blockchain-hit-by-51-attack-leading-to-70k-double-spend}.

\bibitem[Voell(2021)]{etc51}
Zach Voell.
\newblock Ethereum classic hit by third 51\% attack in a month, 2021.
\newblock URL
  \url{https://www.coindesk.com/markets/2020/08/29/ethereum-classic-hit-by-third-51-attack-in-a-month/}.

\bibitem[Porter(2008)]{porter2008five}
Michael~E Porter.
\newblock The five competitive forces that shape strategy.
\newblock \emph{Harvard business review}, 86\penalty0 (1):\penalty0 25--40,
  2008.

\bibitem[Ritala et~al.(2014)Ritala, Golnam, and Wegmann]{ritala2014coopetition}
Paavo Ritala, Arash Golnam, and Alain Wegmann.
\newblock Coopetition-based business models: The case of amazon. com.
\newblock \emph{Industrial marketing management}, 43\penalty0 (2):\penalty0
  236--249, 2014.

\bibitem[Havinga et~al.(2016)Havinga, Hoving, and Swagemakers]{Havinga2016}
Marieke Havinga, Martijn Hoving, and Virgil Swagemakers.
\newblock \emph{Alibaba: A Case Study on Building an International Imperium on
  Information and E-Commerce}, pages 13--32.
\newblock Springer International Publishing, Cham, 2016.
\newblock ISBN 978-3-319-23012-2.
\newblock \doi{10.1007/978-3-319-23012-2_2}.

\bibitem[Zemler(2019)]{zemler2019concepts}
Florian Zemler.
\newblock Concepts for gdpr-compliant processing of personal data on
  blockchain: A literature review.
\newblock \emph{Anwendungen und Konzepte der Wirtschaftsinformatik},
  9:\penalty0 96--107, 2019.

\bibitem[Perboli et~al.(2020)Perboli, Capocasale, and Gotta]{Perboli20201713}
G.~Perboli, V.~Capocasale, and D.~Gotta.
\newblock Blockchain-based transaction management in smart logistics: A
  sawtooth framework.
\newblock In \emph{COMPSAC 2020}, pages 1713--1718, 2020.
\newblock \doi{10.1109/COMPSAC48688.2020.000-8}.

\bibitem[L{\'a}nsk{\`y}(2017)]{lansky2017bitcoin}
Jan L{\'a}nsk{\`y}.
\newblock Bitcoin system.
\newblock \emph{Acta Informatica Pragensia}, 6\penalty0 (1):\penalty0 20--31,
  2017.

\bibitem[Olszewski(2019)]{notDecentralization}
Eric Olszewski.
\newblock Why blockchain matters to enterprise (hint: It’s not because of
  decentralization), 2019.
\newblock URL
  \url{https://medium.com/@eolszewski/why-blockchain-matters-to-enterprise-hint-its-not-because-of-decentralization-8c38674f43c6}.

\end{thebibliography}






\end{document}